# E PLURIBUS ENUM: UNIFYING INTERNATIONAL TELECOMMUNICATIONS NETWORKS AND GOVERNANCE

FINAL CONFERENCE DRAFT: 1 OCTOBER 2001

## Craig McTaggart

Graduate Fellow and Doctoral Candidate
Centre for Innovation Law & Policy
Faculty of Law, University of Toronto
Toronto, Ontario, Canada
t: +1.416.466.4284 • f: +1.416.978.2648
e: <craig.mctaggart@utoronto.ca>
w: <http://www.innovationlaw.org/pages/mctaggartbio.htm>

## Abstract

"ENUM" is an Internet protocol designed to enable one to be reached on an array of different electronic communications devices by means of just one number – a telephone number. ENUM effectively bridges the telephone and Internet worlds by placing telephone numbers from the International Telecommunication Union (ITU) Recommendation E.164 public telecommunication numbering plan into the Internet Domain Name System (DNS) as domain names.

ENUM potentially presents significant public policy issues at both the domestic and international levels. Ultimately, it should not matter whether ENUM is approached as a telecommunications issue or an Internet issue because: (1) they are becoming the same thing technically, and (2) they engage the same global public interests. For the same reasons as apply to traditional telecommunications, and even to the Internet itself, public oversight of ENUM naming, numbering, and addressing resources is justified both by technical necessity and the interests of consumer protection (particularly personal privacy) and competition at higher service layers.

A single, coordinated global DNS domain for at least Tier 0 (the international level) of the ENUM names hierarchy should be designated by public authorities. While the paper deals primarily with Tier 0 issues, it canvasses Tier 1 (domestic) issues because of their relevance for international Internet Tier 0 policy. Many of the technical characteristics and policy considerations relevant at the ENUM Tier 0 and 1 zones are directly applicable to the Internet's IP address space and DNS root (or Tier 0) zone.

Despite the fundamentally international nature of the Internet's logical infrastructure layer, and the purported privatization of administration of its IP address space and the DNS, Internet governance is not yet truly international. Despite assurances to the international community that it would withdraw from this role, the U.S. Department of Commerce has retained significant residual policy authority over the Internet's logical infrastructure. The ENUM policy debate illustrates the need for authoritative international public oversight of public communications network logical infrastructure, including that of traditional telecommunications, the Internet, and ENUM.

## Introduction

"ENUM" is an umbrella term for a series of technical protocols and institutional arrangements which effectively bridge the telephone and Internet worlds. ENUM is designed to enable one to be reached on an array of different electronic communications devices and applications by means of just one identifier - a *telephone number*. ENUM aims to solve some of the biggest problems in the development of Internet Protocol (IP) telephony – namely, different addressing schemes used by public switched telephone network (PSTN) and Internet protocol (IP) terminals, and the lack of public directories of the latter. Because of certain design choices in the current proposal for ENUM, it potentially presents significant public policy issues, at the domestic and international levels. This paper explores some of these issues and their implications for Internet infrastructure policy. It argues that, for the same reasons as apply to traditional telecommunications, and even the Internet itself, public oversight of ENUM naming, numbering, and addressing resources is justified both by technical necessity and the interests of consumer protection (particularly personal privacy) and competition at higher service layers. A single, coordinated global DNS domain for at least Tier 0 (the international level) of the ENUM names hierarchy should thus be designated by public authorities.

Most public policy issues relating to ENUM implementation arise at its Tier 1 and 2 (domestic) levels and are thus the domestic concern of individual countries. In the Internet context, however, since its logical infrastructure is inherently global, the kinds of public policy issues which arise at ENUM Tiers 0 and 1 arise in the Internet's own Tier 0 – the DNS root zone and IP address space – two essential elements of the Internet's logical infrastructure layer. These issues should be thought of as issues of international public policy. While the Internet supports nearly limitless diversity at its higher and lower layers, it demands complete uniqueness of identifiers and absolute adherence to standards and protocols at its *logical infrastructure layer* – if one wants to participate in the global public Internet, that is. The choice to abandon it is always open. While that route is often pursued for particular reasons (e.g., to provide commercial-grade IP telephony), the near-universal interconnectivity and interoperability which the global public Internet offers are hard to leave behind. No one *requires* the use of particular name and number spaces, nor the observance of particular standards and protocols, but choosing not to do so means isolation. The ruthlessly binary nature of computing and computer networking requires strict adherence to relatively narrow but incalculably important sets of rules. *These rules define the Internet*. While they are technical in nature, they are economic, social, political, and effectively legal in consequence.[1]

The field of Internet infrastructure policy is still very young. The purported privatization of the Internet's domain name system (DNS) was the first major event in the development of public policy for the Internet's logical infrastructure. The ongoing development of an international public policy framework for ENUM may well be the second. The question of who should hold what authority (if any) over which elements of the Internet's logical infrastructure is not yet completely settled. *Management responsibility* and *limited administrative authority* over some Internet naming, numbering, and addressing resources have been delegated (perhaps improperly as a matter of U.S. administrative and/or constitutional law – see Section 5(c) below) to the Internet Corporation for Assigned Names and Numbers (ICANN). However, despite assurances that it would divest itself of such power, the U.S. government continues to hold *residual policy authority* over these resources. Considering examples of who holds (or is proposed to hold) these three forms of authority (*management responsibility*, *administrative authority*, and *residual policy authority*) over telephone, Internet, and ENUM naming, numbering, and addressing resources highlights fundamental unresolved international policy issues with respect to the governance of the Internet's logical infrastructure. The existing pattern of Internet governance is not the outcome that many countries expected from the White Paper process.[2] As such, Internet governance is not yet truly international.

Straw-man characterizations of the traditional telecommunications world as state-dominated and rule-bound, and the Internet world as entrepreneur-dominated and unregulated, are no longer true nor helpful. Both have changed, in part due to the impact of one on the other. Neither regime will prevail completely in technical and policy terms, but rather hybrids will emerge. The early popularity of the "*Internet is not telecommunications*" and "*Internet cannot be regulated*" memes only delayed the difficult process of dealing with public policy issues arising from the convergence of traditional telecommunications and the Internet. The principles which underlie policy in each regime need to be examined in an attempt to fashion the right regime for the converged environment, and also for new public communications technologies which resemble neither. To the extent that the dominant trends in computer and communications technology appear to be towards *unified*, *global*, *multiservice* communications platforms, then the associated public policy regimes should similarly strive for *unity*, both in policy and authority.

This paper is primarily about ENUM Tier 0 (or international) policy issues. However, it canvasses Tier 1 domestic issues because of their relevance for international Internet Tier 0 policy. Other papers presented on the same conference panel are presumably concerned with technical and policy considerations at Tiers 1 and 2 in the U.S. context.[3] As a matter of disclosure, my interest in ENUM policy is part of a larger academic interest in Internet infrastructure policy, the subject of my doctoral research in law. I am not currently affiliated with any company with a direct interest in ENUM, nor with any government, nor the International Telecommunication Union (ITU), although I have worked in its General Secretariat before as a consultant on IP telephony policy.

---

[1] This insight is masterfully illustrated in L. Lessig, *Code and Other Laws of Cyberspace* (New York: Basic Books, 1999), albeit with somewhat less recognition of the *international* scope of these issues than would be expected.
[2] United States Department of Commerce, "Management of Internet Names and Addresses" (5 June 1998), <http://www.ntia.doc.gov/ntiahome/domainname/6_5_98dns.htm> (the "White Paper").
[3] J. Hwang et al., "Analyzing ENUM Service and Administration from the Bottom Up: The Addressing System for IP Telephony and Beyond"; R. Cannon, "ENUM Public Policy Considerations"; M. Weiss & H. Kim, "Voice over IP in the Local Exchange: A Case Study". See <http://www.tprc.org/TPRC01/agenda01.htm#enum>.

## 1. **What is ENUM?**

The description of the Internet Engineering Task Force (IETF)[4] working group which developed the ENUM protocol suite provides a good starting-point for understanding ENUM:[5]

> This working group will define a DNS-based architecture and protocols for mapping a telephone number to a set of attributes (e.g. URLs) which can be used to contact a resource associated with that number.
>
> Background:
>
> Telephone numbers now identify many different types of end terminals, supporting many different services and protocols. Telephone numbers are used to identify ordinary phones, fax machines, pagers, data modems, email clients, text terminals for the hearing impaired, etc.
>
> A prospective caller may wish to discover which services and protocols are supported by the terminal named by a given telephone number. The caller may also require more information than just the telephone number to communicate with the terminal.
>
> As an example, certain telephones can receive short email messages. The telephone number is not enough information to be able to send email; the sender must have more information (equivalent to the information in a mailto: URL).
>
> From the callee's perspective, the owner of the telephone number or device may wish to control the information which prospective callers may receive.
>
> The architecture must allow for different service providers competing openly to furnish the directory information required by clients to reach the desired telephone numbers.

ENUM is an acronym – but for precisely what is unclear.[6] In any event, ENUM refers to a series of technical protocols and institutional arrangements which effectively bridge the presently functionally separate worlds of the telephone and the Internet. Richard Shockey, co-chair of the IETF ENUM working group and Senior Technical Industry Liaison with NeuStar, Inc., describes ENUM this way:[7]

> ENUM enables calling users or entities to make a selection from the range of services that are available, especially over the Internet, for communicating with a particular person or entity when the calling user knows only their telephone number. ENUM enables users to access Internet based services and resources from Internet aware telephones, ordinary telephones connected to Internet gateways or proxy services and other Internet connected devices that are limited to numeric keypad data entry, where input is limited to numeric digits. ENUM enables users to specify their preferences for receiving incoming communications (eg, specifying a preference for voicemail messages over live calls or indicating a destination for call forwarding). ENUM will give much improved user control over communications.

---

[4] <http://www.ietf.org/>. The IETF's Website describes it as follows: "The Internet Engineering Task Force (IETF) is a large open international community of network designers, operators, vendors, and researchers concerned with the evolution of the Internet architecture and the smooth operation of the Internet." ("Overview of the IETF" (undated), <http://www.ietf.org/overview.html>).
[5] IETF, Charter, Telephone Number Mapping (enum) Working Group (Last Modified: 31-Jul-01), <http://www.ietf.org/html.charters/enum-charter.html>.
[6] ENUM has been said to stand for: tElephone NUmber Mapping, Electronic NUMber, E-number, and E.164 NUmber Mapping.
[7] R. Shockey, NeuStar, Inc., "Internet-Draft: Frequently Asked Questions About ENUM" (26 July 2000), <http://www.ngi.org/enum/pub/draft-shockey-enum-faq-01.txt>.

ENUM presents two major service opportunities. First, it will provide the first comprehensive way to 'call' an IP telephony-based terminal device, whether from a PSTN phone or another IP phone, regardless of the nature of the underlying long-haul transport network (be it circuit-switched or packet-switched) in between.[8] Second, ENUM provides a way to link (virtually, of course) diverse terminal devices so that they can exchange multimedia messages. Neither ENUM nor its core technology, naming authority pointers (NAPTRs),[9] route or transport the messages – those functions are performed by other protocols and facilities – rather, they identify the available methods for contacting a specific node identified by means of a telephone number, and (optionally) the person's order of preference among these methods at any given time. These might include an Internet email box, an IP phone on a packet-switched network (possibly the Internet), or a fax machine on a circuit-switched network.

To accomplish this, the ENUM protocol (as set out in the remarkably brief RFC 2916[10]) defines a method for converting an ordinary full-length telephone number[11] into a DNS name which can be interpreted by a distributed system of DNS servers to return one or more uniform resource identifiers (URIs) which indicate available communications protocols (and thus devices or applications), the names or addresses associated with them, and the person's preference among which at any given time. The only number which the calling party need know is the called party's telephone number – everything else is automated by a combination of ENUM and other protocols. Further details of ENUM's technical operation, aside from its use of telephone numbers, are not important for present purposes.[12]

The ENUM protocol is an elegantly simple way of converting telephone numbers into domain names. The telephone number is used as a name – not tied to any particular end device or location on any particular network – to identify an individual. In the simplest configuration, a telephone number is entered into a software interface on an Internet-connected device (e.g., a PC). The application converts the string of digits into a domain name by reversing the order, separating each digit with a "." and appending a given second-level domain (SLD) name and a given top-level domain suffix (TLD).

---

[8] For a detailed introduction to the technical, economic, and regulatory issues associated with IP telephony, see International Telecommunication Union, *ITU Internet Reports: IP Telephony* (Geneva: ITU, 2000), of which Chapter 4, "Regulatory Aspects of IP Telephony," was written by the author while working as a consultant in the ITU's Strategy & Policy Unit (SPU) in 2000.
[9] RFC 2915, M. Mealling, Network Solutions, Inc. and R. Daniel, DATAFUSION Inc., "The Naming Authority Pointer (NAPTR) DNS Resource Record" (September 2000), <http://www.ietf.org/rfc/rfc2915.txt>.
[10] RFC 2916, P. Fältström, Cisco Systems Inc., "E.164 number and DNS" (September 2000), <http://www.ietf.org/rfc/rfc2916.txt>.
[11] Including all country codes (for international calls) and area codes (for calls within North America, for example), but not dialing codes, such as the leading "1" used in North America as a flag for switches to identify long distance calls.
[12] Web pages offering comprehensive links to ENUM-related technical and policy material are available at <http://www.cybertelecom.org/enum.htm> (Washington Internet Project), <http://www.itu.int/osg/spu/infocom/enum/index.html> (ITU), and <http://www.ngi.org/enum/> (Centre for Next Generation Internet).

Thus the full-length telephone number "+1 (703) 845-1010" (the reception desk of the Hilton Hotel at Mark Centre, Alexandria, Virginia, USA) becomes "0.1.0.1.5.4.8.3.0.7.1.<sld>.<tld>". This Uniform Resource Locator (URL) is functionally no different from "tprc.org" – it simply involves more domains to the left of the TLD (each dot demarcates a different domain).

## 2. **E.164 Numbers**

To the extent that ENUM might appear to be merely a new way of using the Internet's existing DNS, any policy issues relating to it might be assumed to be *Internet* policy issues. Indeed, the choice of telephone numbers at first appears arbitrary – any string of ASCII[13] characters could have been used instead, such as a domain name, in which case the policy issues at stake would certainly be limited to the Internet (but would be no less international public policy issues). However, RFC 2916 chooses a particular set of characters to use as the initial 'hook' into the ENUM system – "E.164" *telephone numbers*. Patrik Fältström's proposal calls for the "transformation of E.164 numbers into DNS names.[14] E.164 refers to the ITU's "international public telecommunication numbering plan"[15] (and other associated standards). The E.164 plan is a kind of 'master list' of telephone numbering plans in use around the world (identified by Country Code (CC)) and, consequently, the agencies which have authority over them. Choosing to use telephone numbers *as telephone numbers* (as distinct from how they might be used at, say, a video store for membership indexing) has policy consequences (of which Fältström, a Swede, was quite aware).

Country codes (and other global codes which are not discussed herein) are allocated and managed by the Director of the ITU Telecommunication Standardization Bureau (TSB) and ITU Member States. The role of the TSB Director and the responsibilities of the ITU Member States with respect to the numbering resources in question are defined in a World Telecommunication Standardization Assembly (WTSA) resolution titled: "Resolution 20 – Procedures for allocation and management of international numbering resources."[16] At the top level, the telephone numbering space is only an administrative hierarchy (as opposed to an operational hierarchy like the DNS[17]). A CC must, for technical reasons, only be used by one country (or other political entity, or group thereof), and within each country, numbers falling under each CC must similarly be managed to ensure non-conflicting use. The countries of the world have agreed that the top level zones of international telecommunications numbering resources should be neutrally and transparently managed by the TSB. However, control over each country's assigned E.164 resources lies entirely within that country by the associated sovereign.

---

[13] American Standard Code for Information Interchange.
[14] Note 10 above, at 1: "Through transformation of E.164 numbers into DNS names and the use of existing DNS services like delegation through NS records, and use of NAPTR records in DNS, one can look up what services are available for a specific domain name in a decentralized way with distributed management of the different levels in the lookup process." (references omitted).
[15] ITU-T Recommendation E.164, "The international public telecommunication numbering plan (05/97)", <http://www.itu.int/itudoc/itu-t/rec/e/e164.html>.
[16] ITU, World Telecommunication Standardization Assembly (WTSA), Montreal, 2000, "Resolution 20 - Procedures for allocation and management of international numbering resources", <http://www.itu.int/itudoc/itu-t/wtsa-res/res20.html>.
[17] While the political motivation for the document is plain, A.M. Rutkowski's Internet-draft, "Development and Basis of International Telephone Numbering" (21 March 2001), <http://www1.ietf.org/internet-drafts/draft-rutkowski-enum-basis-00.txt> provides a detailed history of the E.164 plan and its predecessors.

**3.      Zones of Authority**

The Internet DNS is a system of hierarchical, distributed databases. Authoritative lists of names within each domain must be maintained to prevent conflicting use, and thus to ensure the integrity of the DNS as a whole. In the example of "tprc.org", three such lists are implied, from right to left: that of what TLDs are in the 'root zone' (the suppressed trailing dot after "org"[18]), what SLDs are in the "org" TLD, and what hosts are in the "tprc" SLD. Many lists like these make up the DNS databases. The DNS protocol mechanisms effect the mapping of domain names into Internet IP addresses, which identify individual Internet-host interfaces.[19] These lists are kept on name servers in the form of zones – particular sections of the overall DNS 'tree'.[20]

Much of the day-to-day work of ICANN since its creation has consisted of formalizing the legal framework governing the registries and registrars of domains and domain names in this tree. As explained in Section 5(c) below, ICANN has been given limited administrative authority over the DNS. Management responsibility lies with various registrars (e.g., the registrar of .com is VeriSign Global Registry Solutions, a subsidiary of VeriSign, Inc., which also owns Network Solutions, Inc. – the original .com registrar) while residual policy authority rests with the U.S. Department of Commerce (DoC). Transforming E.164 telephone numbers into domain names implies the identification of who holds management responsibility, administrative authority, and residual policy authority over each domain in an ENUM name such as "0.1.0.1.5.4.8.3.0.7.1.<sld>.<tld>".

Similarly, the fundamental question in Internet governance is that of who should hold what authority or responsibility (if any) over the suppressed dot at the far right of every Internet domain name and email address (the root of the Internet naming tree) and the IP address space to which DNS protocols map domain names and email addresses. These resources are referred to in this paper as 'Tier 0' of the Internet's naming, numbering, and addressing spaces. The concept of 'tiers' is key to understanding ENUM, as explained below.

---

[18] RFC 1034, P. Mockapetris, ISI, "Domain Names – Concepts and Facilities" (November 1987), <http://www.ietf.org/rfc/rfc1034.txt> at Section 3.1.
[19] Internet traffic is *routed* by reference to IP *addresses*, not domain *names*. The meanings of these terms are explained in H. Rood, "What's in a name, what's in a number: some characteristics of identifiers on electronic networks" (2000) 24 Telecommunications Policy 533 at 538: "A name identifies a person, a device, an organisation, a role or a place; [a]n address identifies a location of a device on an electronic network; and [a] route identifies a path to reach a certain address."
[20] Note 18 above, at Section 2.4: "NAME SERVERS are server programs which hold information about the domain tree's structure and set information. A name server may cache structure or set information about any part of the domain tree, but in general a particular name server has complete information about a subset of the domain space, and pointers to other name servers that can be used to lead to information from any part of the domain tree. Name servers know the parts of the domain tree for which they have complete information; a name server is said to be an AUTHORITY for these parts of the name space. Authoritative information is organized into units called ZONEs, and these zones can be automatically distributed to the name servers which provide redundant service for the data in a zone."

**4.**          **Should There Be a Single, Coordinated Global DNS Domain for ENUM?**

*(a)*          *Designated or multiple zones at Tier 0?*

While authority over each domain in the ENUM name "0.1.0.1.5.4.8.3.0.7.1.<sld>.<tld>" must be analyzed separately, the domains naturally group into tiers, making the analytical process simpler. While the concept of tiers does not appear in RFC 2916, it is nearly-universally used in ENUM discourse. ENUM 'Tier 0' consists of the "<sld>.<tld>" combination in the above example, that is, the authoritative list of international CC zones in the global ENUM namespace. CCs comprise the first one to three digits to the left of "<sld>". In the example, "+1" is the CC, and with the Tier 0 information added, becomes "1.<sld>.<tld>" – 'Tier 1' in the ENUM schema. Of course, "+1" is a special case because it is shared by a group of countries in North America and the Caribbean.[21] Replacing "+1" with "+678" (Vanuatu), and taking into account the structure of the national numbering plan within "+678", the example becomes: "9.4.7.6.3.8.7.6.<sld>.<tld>" (the fax machine at the Hotel Santo in Luganville, Espiritu Santo Island, Vanuatu). In this case, "8.7.6.<sld>.<tld>" is a single Tier 1 zone. 'Tier 2' refers to the full-length ENUM domain name, which mirrors the full-length E.164 number, but in reverse, with the "+" stripped off, with dots separating the digits, and the SLD and TLD tacked on the end – that is, "9.4.7.6.3.8.7.6.<sld>.<tld>".

The initial question in the ENUM policy debate (and the focus of the first half of this paper) is whether there should be only one Tier 0 list of Tier 1 zones and authoritative name servers for each CC zone. If Tier 0 is so designated, then the decision of which TLD is to be used to house ENUM names must be made.[22] Alternatively, if Tier 0 is not designated, then there may presumably be multiple lists (which can technically be placed under any "<sld>.<tld>" combination) serving the same function. VeriSign, Inc. and its representatives have from time to time stated that it considers itself at liberty to set up an ENUM name space under the TLD and SLD of its choosing and start offering Tier 2 services (that is, the actual conversion of E.164 phone numbers into Uniform Resource Identifiers (URIs)). This, more than anything else, has gotten the attention of the ITU-T and national numbering plan administrators because it raises the possibility that telephone number allocations in the ENUM name space might diverge from those in the 'official' E.164 numbering plan space. Their concerns are summarized in the following passage from the Autorité de Régulation des Télécommunications (ART), the numbering plan authority in France:[23]

---

[21] The North American Numbering Council (NANC) administers "+1" under the North American Numbering Plan (NANP), <http://www.fcc.gov/ccb/Nanc/>. Technical management is performed by the North American Numbering Plan Administrator (NANPA) (currently NeuStar, Inc.). Special arrangements will have to be made among the countries within 'World Zone 1' regarding Tier 1. Within "+1", numbering plan areas (NPAs) are assigned to specific areas (including provinces, states, cities, and countries) within the larger area. Within World Zone 1, it may be useful to think of the top-level zone as 'Tier 1a', while a group of NPAs relating to a particular country might be referred to as 'Tier 1b', and specific NPAs within each country as 'Tier 1c'. For the sake of simplicity, these divisions are not used in this paper.

[22] Due to space constraints, this issue is not addressed in detail in this paper, but see Section 4(e) below.

> Primarily, ENUM raises the question of the coherence between E.164 telephone numbers and domain names. If the general public is to use services made possible by ENUM, it appears that perfect coherence will have to be guaranteed in order to protect the main advantage of E.164 numbering, which is the use of a system already widely used and accepted by the public.
>
> Special attention should be paid to this question, for the following reasons:
>
> - Management of ENUM domain names which is not coordinated with that of E.164 numbers could result in the creation of ENUM subdomains which do not correspond to the country codes assigned by the ITU.
>
> - Similarly, management of ENUM domain names which is not coordinated with that of E.164 numbers could result in the assignment of ENUM domain names which do not respect the numbering plan matching a given country code.
>
> - Finally, poor correspondence between the assigned ENUM domain names and E.164 numbers could cause incoherence between the recipient of a telephone number and the recipient of the corresponding ENUM domain name.

Maintaining "perfect coherence" between E.164 numbers and their equivalent ENUM names implies that there be only one 'official' ENUM name space – often called the 'Golden Tree'. Apart from the policy reasons noted by ART, others of which are discussed below, there are simple technical reasons why this is not only desirable, but necessary.

### (b) The technical necessity of uniqueness

One of the most controversial issues in Internet governance is, surprisingly, the question of whether "perfect coherence" is required in the Internet's naming, numbering, and addressing spaces. That the controversy persists is somewhat odd, given the preponderance of technical opinion that such uniqueness is absolutely essential – a *sine qua non* of public internetworking. Indeed, the exact same considerations apply to all three of telephone networks, the Internet, and ENUM. With respect to the Internet, the Internet Architecture Board (IAB)[24] explains:[25]

> To remain a global network, the Internet requires the existence of a globally unique public name space. The DNS name space is a hierarchical name space derived from a single, globally unique root. This is a technical constraint inherent in the design of the DNS. Therefore it is not technically feasible for there to be more than one root in the public DNS. That one root must be supported by a set of coordinated root servers administered by a unique naming authority.

---

[23] France, Autorité de Régulation des Télécommunications (ART), "The principles and conditions of implementation of the ENUM protocol in France – Public consultation: 23 May 2001 – 12 June 2001" (English version), <http://www.art-telecom.fr/publications/cp-enum-gb.htm>.
[24] <http://www.iab.org/iab/>. Its Website describes it as follows: "The Internet Architecture Board (IAB) is a technical advisory group of the Internet Society." ("IAB Overview" (3 November 2000), <http://www.iab.org/iab/overview.html>). The Internet Society (ISOC) (<http://www.isoc.org/>) describes itself as "a professional membership society with more than 150 organizational and 6,000 individual members in over 100 countries. It provides leadership in addressing issues that confront the future of the Internet, and is the organization home for the groups responsible for Internet infrastructure standards, including the Internet Engineering Task Force (IETF) and the Internet Architecture Board (IAB).".
[25] RFC 2826, Internet Architecture Board, "IAB Technical Comment on the Unique DNS Root" (May 2000), <http://www.ietf.org/rfc/rfc2826.txt> at 1.

> Put simply, deploying multiple public DNS roots would raise a very strong possibility that users of different ISPs who click on the same link on a web page could end up at different destinations, against the will of the web page designers.
>
> This does not preclude private networks from operating their own private name spaces, but if they wish to make use of names uniquely defined for the global Internet, they have to fetch that information from the global DNS naming hierarchy, and in particular from the coordinated root servers of the global DNS naming hierarchy.

Since ENUM is a DNS-based system, the exact same considerations apply. As identified by ART, one risk inherent in multiple Tier 0 lists (and associated databases and servers) is that one number may be assigned to different people in the E.164 hierarchy and the ENUM hierarchy, whether accidentally or intentionally. The importance of coherence in the assignment of domain names is also well explained by Stuart Lynn, President of ICANN:[26]

> The DNS is a globally distributed database of domain name (and other) information. One of its core design goals is that it reliably provides the same answers to the same queries from any source on the public Internet, thereby supporting predictable routing of Internet communications. Achievement of that design goal requires a globally unique public name space derived from a single, globally unique DNS root.

Uniqueness is also required in Internet IP addressing and telephone network numbering for the exact same reasons.[27]

### (c)     What about 'unofficial' ENUM name spaces?

The technical necessity of a single, authoritative top-level ENUM root zone may be one thing, but saying that *there can be only one* such zone (and thus name space) is another. VeriSign's Vice-President of Internet Strategy, Anthony Rutkowski, asserts that multiple ENUM roots should not only be 'permitted,' but that they may be the only possible way to implement ENUM at the global level.[28] VeriSign's proposal for the ENUM name space appears to be that competing firms would populate and maintain their own Tiers 0, 1 and 2 databases by entering into commercial agreements with all necessary parties. The possibility that a private firm might proceed with such a plan (even if just as a 'trial') before an 'official' Tier 0 regime can be designated through ITU-T presents the question of what public authorities could do about such 'unofficial' implementations, if they did become widely used.

First, it is worth bearing in mind these further comments by ICANN President Stuart Lynn, with reference to the Internet name space more generally:[29]

> [ICANN's mandate to preserve stability of the DNS does not preclude] experimentation done in a manner that does not threaten the stability of name resolution in the authoritative DNS. Responsible experimentation is essential to the vitality of the Internet.

---

[26] S. Lynn, President, ICANN, "ICP-3: A Unique, Authoritative Root for the DNS" (9 July 2001), <http://www.icann.org/icp/icp-3.htm>.

[27] See A. Andeen & J.L. King, "Addressing and the Future of Communications Competition: Lessons from Telephony and the Internet," in B. Kahin & J.H. Keller, eds., *Coordinating the Internet* (Cambridge, MA: MIT Press, 1997) 208.

[28] See, for example, A. Rutkowski, "the ENUM golden tree: the quest for a universal communications identifier" (2001) 3 info 97, <http://www.ngi.org/enum/pub/info_rutkowski.pdf>.

[29] Note 26 above.

> Nor does it preclude the ultimate introduction of new architectures that may ultimately obviate the need for a unique, authoritative root. But the translation of experiments into production and the introduction of new architectures require community-based approaches, and are not compatible with individual efforts to gain proprietary advantage.

Quite apart from the question of the authority of ICANN to prohibit 'alternate' DNS roots, or that of the ITU to prohibit alternate ENUM roots (which neither likely have on their own), is the practical question of whether alternate roots would be likely to succeed against 'official'[30] roots at all. For the same reason that a single authoritative root is necessary from a technical standpoint, it is also preferable from a business standpoint for users of all kinds who rely on the stability of the DNS for their own purposes (as opposed to firms wanting to enter the name and number assignment business). Firms of the former type will naturally prefer the lists that 'everyone' uses. Jonathan Weinberg explains that the DNS root which ICANN now administers is considered 'the' public Internet root today, even though alternative roots have been available for years for those who wish to refer to them:[31]

> Very few Internet users … look to alternative root servers. The vast majority rely on the single set of authoritative root servers, historically supervised by Jon Postel, that have achieved canonical status.

Indeed, Rutkowski himself believes that it would probably not be necessary to prohibit alternative ENUM roots because the official one(s) would trump them in any case:[32]

> The Followers [his term of derision for proponents of the single, authoritative root model] – while casually suggesting that others can offer competitive ENUM services outside the Golden Tree – know full well that the existence of a government-designated zone will significantly harm, if not exclude competitors from the marketplace.

This is somewhat surprising to read, given both Rutkowski's otherwise resolute faith in the superiority of marketplace solutions at all tiers and the fact that his company has recently been forced to abandon its historical claims to anti-trust immunity relating to its activities under the Cooperative Agreement between it and the DoC.[33] Being government-designated, both the root server and the gTLD registries managed by VeriSign subsidiary NSI would presumably have the same negative consequences on the prospects for competition in, at least, the gTLD name space – perhaps partially explaining why VeriSign thought *US$21 billion* a fair price for NSI in March 2000.[34]

---

[30] It is worth noting that the administrators of the many networks which make up the public Internet are not under any *legal* obligation to point their name servers to the root servers administered by ICANN, thus alternative (or 'unofficial') systems are at least legally possible.
[31] J. Weinberg, "ICANN and the Problem of Legitimacy" (2000) 50 Duke L.J. 187, <http://www.law.wayne.edu/weinberg/legitimacy.PDF> at 198.
[32] Note 28 above, at 99.
[33] See note 97 below.
[34] "VeriSign Nabs NetSol for $21 Bil" *Wired News* (7 March 2000), <http://www.wired.com/news/business/0,1367,34784,00.html>.

In Rutkowski's view, the question of unofficial ENUM name spaces should be understood in anti-trust (or competition law) terms. A sense of resignation to the inevitability of a designated Tier 0 is evident in the following statement:[35]

> This is an informational Internet Draft containing a proposed technical and administrative framework and elements for ENUM neutrality among implementations.
>
> Such a framework is important to assure that if some special public governmental or intergovernmental arrangements are sought for particular ENUM offerings in the marketplace, such actions: 1) avoid prejudice to commercial ENUM offerings employing other element options, and 2) provide for fair and non-discriminatory access by other commercial competitive ENUM providers to publicly supported administrative and information resources.

Here Rutkowski appears to argue (on behalf of VeriSign, Inc.), that if a single authoritative ENUM root is designated, then despite there being only one registry and registrar at the Tier 0 (and presumably Tier 1) levels, other firms should have a right of *access* to the underlying data which they would need to construct their own databases at those layers.

This argument acknowledges that operating a competitive ENUM system is not as simple as simply replicating existing E.164 numbers in a given DNS domain. While it is certainly feasible, assuming that one could access comprehensive, up-to-date information on E.164 telephone number assignments all over the world (or some sub-set thereof), the difficulty would be keeping that information up-to-date and validating the identity and authority of new and existing customers to add or change information in the system. Other changes, such as area code overlays and splits, could also threaten the integrity of a competing ENUM Tier 0 and 1 registry if its operator does not have real-time access to updated information. However, if a particular national numbering plan authority intends that there be only one authoritative Tier 1 list (and associated database and server system) applicable to a country's telephone numbers, then it could 'starve' would-be alternative operators by not making such data available. Would such a policy be anti-competitive (under North American competition law principles, for the sake of illustration), as Rutkowski argues?[36]

First is the elementary point that competition laws do not apply to governments *acting as governments* (as opposed to as a firm in a market). If it did, then government monopolies over the printing of money, collection of taxes, and operation of courts could similarly be branded anti-competitive. In countries in which the 'rule of law' is a fundamental constitutional principle, governments enjoy wide latitude to do what they consider right, within certain limits. Competition law is not, however, a source of such limits.

---

[35] A.M. Rutkowski, VeriSign, Inc., "Framework for ENUM Neutrality" (29 August 2001), <http://www.ngi.org/enum/pub/draft-rutkowski-enum-neutrality-00.txt>.
[36] Note 28 above, at 99: "There are major anti-competitive implications of the Golden Tree."

Second, the rationale underlying telephone number administration in competitive markets illustrates why designating a Tier 1 list would be explicitly *pro-competitive* – that is, in the interests of competition at higher service layers. Control over numbering resources used in public networks (that is, outside of 'private dialling plans' or internal networks) is widely recognized as a potential source of anti-competitive activity in higher-layer service markets, particularly where one firm holds a dominant position in one or more of them.

Even New Zealand, renowned for its 'hands-off' telecommunications regulatory regime, has mandated a neutral industry-based means of administering telephone numbers, in the interest of competition itself.[37] While administrative authority is exercised through the industry group, residual policy authority is held by the government. The same type of regime prevails in Canada, where the Canadian Numbering Administration Consortium Inc. (CNAC) (holder of administrative authority) engages the Canadian Numbering Administrator (CNA)[38] to perform technical management functions at its direction. CNAC is ultimately subject to the residual policy authority of the Canadian Radio-television and Telecommunications Commission (CRTC).[39] It is likely that many countries with pro-competitive telecommunications regulatory regimes will consider imposing a similar structure at ENUM Tier 1, in the interests of both competition at Tier 2 and subscriber privacy. Implicit in this approach, however, is the assertion that ENUM concerns telecommunications, and is not simply an Internet directory service. The next section addresses this important interpretive issue.

### (d)   Is ENUM a telecommunications policy issue or an Internet policy issue?

Another of Rutkowski's arguments against a single, designated ENUM root zone is that ENUM has nothing to do with the international public telecommunications numbering plan, but rather is merely an *Internet-based directory service*. Following his comment that designating official ENUM zones is anti-competitive, Rutkowski continues:[40]

---

[37] See New Zealand, Ministerial Inquiry into Telecommunications, "Final Report" (27 September 2000), <http://www.med.govt.nz/teleinquiry/reports/final/final-07.html#P1020_190362> at section 7.4. The New Zealand government is in the process of introducing new telecommunications legislation based on the following premises (which are rather different from those underlying the previous competition-law-only regime, which is widely thought to have been an abject failure): "1. There should be a single regulatory framework covering all electronic communications services. 2. The existing regulatory regime is not best suited to achieving the Government's objective for electronic communications. To meet the objective, light-handed industry specific regulation (the Electronic Communications Act) is required." (at "Recommendations").
[38] Currently SAIC Canada. See <http://www.cnac.ca/>.
[39] Further, access to subscriber listing information, which is likely what Rutkowski means by "publicly supported administrative and information resources" is often regulated, primarily to protect subscriber privacy. In Canada, for instance, directory data is made conditionally available monthly by law to "independent telephone directory publishers" (Telecom Decision CRTC 95-3, *Provision of Directory Database Information and Real-Time Access to Directory Assistance Databases* (8 March 1995), <http://www.crtc.gc.ca/archive/eng/Decisions/1995/DT95-3.htm>, as amended by Order-in-Council P.C. 1996-1001 (25 June 1996), while *real-time* data (which ENUM services would almost certainly require) is only made conditionally available "on condition that it be used solely for the purpose of providing directory assistance" (Telecom Decision CRTC 97-8, *Local Competition* (1 May 1997), <http://www.crtc.gc.ca/archive/eng/Decisions/1997/DT97-8.htm> at para. 233). ENUM services would likely not be considered directory assistance because, as that term is ordinarily understood in North America, it implies that the requester does not know the phone number of the person he or she wants to reach. ENUM, in contrast, is based on the use of known phone numbers. ENUM service providers would thus probably have a legal right of access to neither monthly or real-time subscriber listing information.

> What tend to get swept under the carpet are fundamental issues like whether a co-ordinated government movement should meddle in a nascent market, based on some new internet protocol. Just because this is an internet directory service based on telephone numbers doesn't seem a good enough reason.

While he has puzzlingly contradicted this sentiment elsewhere by saying that ENUM is *not* a directory service,[41] his argument appears to be that an Internet directory service (even if "based on telephone numbers") is beyond the competence of telecommunications regulatory authorities. However, his concomitant recognition that ENUM is "based on telephone numbers" is telling.

As explained above, the choice to use E.164 telephone numbers for ENUM was anything but arbitrary. While it is often forgotten, the original motivation for ENUM (and other such efforts) was primarily to find a way to enable PSTN subscribers to dial an IP phone – currently only the opposite is possible outside of private network environments. The other options, such as IP addresses, cannot easily be dialed on a standard telephone, and even if they could, mass market acceptance might be hard to come by. Telephone numbers, on the other hand, are familiar, easy to remember, easy to dial, and associated with a system thought to be stable, reliable, and relatively secure – the telephone system. Having made the decision to use E.164 resources, the Internet community (with which VeriSign currently appears to be at odds, as has often been the case in the past) must be prepared to abide by the rules applicable to telephone numbers, at the domestic and international levels. Indeed, in RFC 2916, Fältström explicitly recognizes this:

> Names within this [ENUM top-level] zone are to be delegated to parties according to the ITU recommendation E.164. The names allocated should be hierarchic in accordance with ITU Recommendation E.164, and the codes should assigned in accordance with that Recommendation.

Not only is there consensus within the Internet technical community on both the technical and policy necessity of a single, coordinated ENUM Tier 0, but the relevant government agencies in Sweden,[42] the United Kingdom,[43] and France[44] have expressed initial support for the Golden Tree model. Even the industry advisory group studying the issue in the U.S. has recommended that: "[t]he US should participate in a global implementation of ENUM rooted in e164.arpa per RFC 2916."[45] However, the choice of

---

[40] Note 28 above, at 99-100.
[41] In A.M. Rutkowski, "ENUM – networking's new *glueball* infrastructure," presented at 3rd European ISP Forum, Rome, Italy (2-5 October 2000), <http://www.ngi.org/enum/pub/enum_presentation.zip>, Rutkowski asserts that ENUM is: "Not a directory service, but a core infrastructure for interoperating with all other network infrastructure" (at slide 4) (emphasis in original).
[42] Sweden, Post & Telestryrelsen, "ENUM - functions that maps telephone numbers to Internet based addresses - a description and the possible introduction to Sweden" (23 March 2001), <http://www.enum.org/information/files/enum_summary.pdf>
[43] United Kingdom, Department of Trade and Industry (DTI), Communications and Information Industries Directorate, "DTI ENUM Workshop 5 June 2001-06-06: Summary", <http://www.dti.gov.uk/cii/regulatory/enum/>.
[44] See note 23 above.
[45] "Report of the Department of State ITAC-T Advisory Committee Study Group A Ad Hoc on ENUM" (6 July 2001), <http://www.nominum.com/ENUM/2001_07_06-ENUM-Report-Department-of-State-Final.html> at section 8.2.

*which* TLD (".arpa" or otherwise) ENUM should be rooted in remains unresolved. Following on its heels is the daunting task of sorting out who should hold what authority over the many Tier 1 and 2 zones in the global ENUM system. The contentious TLD issue is renewing interest among countries in the question of control over the Internet's Tier 0 zones themselves.

### (e) Which TLD?

A complex series of steps has been taken by the IAB to put itself in control of "e164.arpa",[46] so that it can negotiate with ITU-T over administrative authority and management responsibilities for ENUM Tier 0. The details of this negotiation, which continue at time of writing, are, due to space constraints, beyond the scope of this paper. The current IAB proposal, however, appears to be that it would hold administrative authority for "e164.arpa", while RIPE NCC,[47] which is also the Regional Internet Registry (RIR) for IP addresses in Europe, would have management responsibility for "e164.arpa", under the IAB's direction. The IAB's explanation for the choice of ".arpa" as the ENUM TLD is that it is the one existing Internet TLD which serves an explicit infrastructure function, and is 'hardened' with the kind of security and reliability features that an infrastructure domain like ENUM's will need[48] – a claim that arguably does not bear much scrutiny.[49] Whether ITU-T[50] Study Group 2 (SG2)[51] will endorse this model remains to be seen, but not surprisingly there has already been rejection by several ITU Member States of the idea that the IAB or ICANN should have any administrative control over ENUM Tier 0.

In August 2001, as another SG2 meeting approached, John Klensin wrote a liaison document from the IAB, on behalf of the IETF, to SG2.[52] A sense of exasperation with the slow pace of defining initial operational procedures, and with the dismaying prospect of 'unofficial' ENUM roots being launched, is evident in the tone of the document. Klensin affirms the IAB and IETF's commitment to ITU authority over the E.164 numbers which are placed into the DNS:[53]

---

[46] See "Chairman's Report of the ITU ENUM Workshop", ITU, Geneva - 17 January 2001", <http://www.itu.int/osg/spu/infocom/enum/workshopjan01/report-jan17-2001.html>, "Annex 7: ENUM Issues", <http://www.itu.int/osg/spu/infocom/enum/workshopjan01/annex8-responsibilitiesfore164arpa.html>.
[47] Réseaux IP Européens (RIPE) Network Coordination Centre (NCC), <http://www.ripe.net/ripencc/index.html>.
[48] See John Klensin, for the IAB, "IAB Statement on Infrastructure Domain and Subdomains" (undated), <http://www.iab.org/iab/DOCUMENTS/iab-arpa-stmt.txt>.
[49] For example, ".arpa" contravenes the IETF's 'best current practice' recommendation that the DNS root servers not provide services other than root name service (RFC 2870, R. Bush, *et al.*, "Root Name Server Operational Requirements" (June 2000), <http://www.ietf.org/rfc/rfc2870.txt>), and the ".arpa" name servers are not nearly as globally distributed as they would need to be to carry out a global infrastructure function (see R. Shaw, ITU, "Global ENUM Implementation" presented at DTI ENUM Workshop, London, England (5 June 2001), <http://www.itu.int/osg/spu/infocom/enum/dtijune501/dti-june-5-2001-1.PPT> at slide 15 (showing that 8 of 9 are physically located within the U.S., the 9$^{th}$ in Stockholm, Sweden).
[50] <http://www.itu.int/ITU-T/info/itu-t/index.html>. ITU-T describes itself as follows: "The ITU-T mission is to ensure an efficient and on-time production of high quality standards covering all fields of telecommunications except radio aspects. Standardization work is carried out by 14 study groups in which representatives of the ITU-T membership develop Recommendations for the various fields of international telecommunications on the basis of the study of Questions (i.e. areas for study)." ("About ITU-T" (undated), <http://www.itu.int/ITU-T/info/itu-t/about.html>).
[51] <http://www.itu.int/ITU-T/studygroups/com02/index.html>. SG2 focuses on operational aspects of telecommunications service provision, networks, and performance, including naming, numbering, and addressing requirements and resource assignment. ("About ITU-T," *ibid.*)
[52] Internet Architecture Board (IAB) on behalf of the IETF, "Reflections on risks of, and barriers to, ENUM deployment" (August 2001), <http://www.iab.org/iab/DOCUMENTS/iab-sg2-liaison-3.txt>.

> As with the E.164 system itself (and the DNS more broadly), if users are to have confidence that a particular number will reach the intended party or resource, independent of who is asking the question or where they are asking from, it is necessary to have only one way to access and interpret that number. […]
>
> While there have been a number of proposals for independent schemes with no central authority or coordination, *those schemes either deny the obvious linkage between ENUM identifiers and E.164 numbers (claiming that the former just "look like" telephone numbers and are easy to remember, but that they are completely independent and no one will confuse the two)* or assume a different structure in the Domain Name System than it actually uses, based on coordinated national databases, and that would add little or no value for the user of the anticipated services.

Behind the issue of whether the IAB or ICANN should have any control over Tier 0 is the reality that ".arpa", like all other public Internet TLDs, is subject to the residual control of the U.S. government. Klensin explains the history:[54]

> The current relationship with the US Government is much the same [as during the ARPANET period]: the registry for the domain is the IANA, operating under IAB supervision. The Defense Department has formally relinquished any claims on it that they might have had (and that few, even on their staff, believed that they did have). And the domain itself has the same relationship with the US Department of Commerce that any other TLD, including country code TLDs and TLDs which are not country-specific such as .INT, has: in principle, the US Government could order the root operator to make changes against the will of the users of that domain.
>
> In the hope of avoiding future confusion and to further identify the infrastructure purpose of the domain, *we have begun to identify the domain name as an acronym for "Address and Routing Parameter Area"*. Of course, this does not change any of the underlying relationships, which are described above.

We will return to this key issue after a brief note regarding instant messaging names and databases.

### (f)           *AOL Instant Messenger "Names and Presence Database"*

It is worth noting briefly that even within the information services realm, regulatory action is sometimes necessary to address potential anti-competitive behaviour relating to identifiers. In the decision following its review of the AOL/Time Warner merger, the U.S. Federal Communications Commission (FCC), in a classic telecommunications policy analysis, found as follows:[55]

> We conclude the market in text-based instant messaging is characterized by strong "network effects," i.e., a service's value increases substantially with the addition of new users with whom other users can communicate, and that AOL, by any measure described in the record, is the dominant IM provider in America. We further find AOL has consistently resisted interoperability with other non-licensed IM providers. AOL's market dominance in text-based messaging, coupled with the network effects and its resistance to interoperability, establishes a very high barrier to entry for competitors that contravenes the public interest in open and interoperable communications systems, the development of the Internet, consumer choice, competition and innovation. *We also find that a Names and Presence Database ("NPD") is currently an essential input for the development and deployment of many, if not most, future high-speed Internet-based services that rely on real-time delivery and interaction.*

---

[53] *Ibid*. (emphasis added).
[54] *Ibid*. (emphasis added).
[55] United States, Federal Communications Commission, "In the Matter of Applications for Consent to the Transfer of Control of Licenses and Section 214 Authorizations by Time Warner Inc. and America Online, Inc., Transferors, to AOL Time Warner Inc., Transferee" (22 January 2001), <http://www.fcc.gov/Bureaus/Cable/Orders/2001/fcc01012.txt> at para. 59 (emphasis added).

The Commission later elaborates that "[t]he NPD is more than simply a customer list. It is a working part of an electronic communications network for persons who have requested participation in the network and actually use it to exchange communications in real time with other users."[56] To ensure that the public interest in the U.S. is served by interoperability among NPD-based services, the FCC imposed a condition on future versions of AOL's IM service which relate directly to the NPD:[57]

> AOL Time Warner may not offer an AIHS ["Advanced IM-based High-Speed Service"] application that includes the transmission and reception, utilizing an NPD over the Internet Protocol path of AOL Time Warner broadband facilities, of one- or two-way streaming video communication using NPD protocols including live images or tape that are new features, functions, and enhancements beyond those offered in current offerings such as AIM 4.3 or ICQ 2000b, unless and until AOL Time Warner has successfully demonstrated it has complied with one of the following grounds for relief.[58]

The FCC's regulatory treatment of AOL's "Names and Presence Database" demonstrates that even apart from traditional telecommunications numbering authority, regulators may find reason to impose rules on public communications networks when the public interest requires it.

### 5. What Do ENUM Policy Issues Tell Us About Internet Policy Issues?

#### (a) *The unavoidable need for uniqueness and authority*

The technical reasons for a single, coordinated global ENUM DNS domain are equally applicable to the DNS itself. While there has always been a certain amount of minority opinion about this in the Internet community,[59] the discussions of the technical requirements of ENUM above demonstrate the parallel necessity for uniqueness throughout the DNS and IP address spaces. Even David Post, whose early writing enthusiastically embraced forms of Internet coordination *other* than 'top-down' authority,[60] subsequently acknowledged this fact with respect to the DNS:[61]

---

[56] *Ibid.*, at para. 68.
[57] *Ibid.*, at para. 121.
[58] The three options are: "**Option One**. AOL Time Warner may file a petition demonstrating that it has implemented a standard for server-to-server interoperability of NPD-based services that has been promulgated by the IETF or a widely recognized standard-setting body that is recognized as complying with National Institute of Standards and Technology or International Organization for Standardization requirements for a standard setting body." "**Option Two**. AOL may file a petition demonstrating that it has entered into written contracts providing for server-to-server interoperability with significant, unaffiliated, actual or potential competing providers of NPD-based services offered to the public." "**Option Three**. AOL Time Warner may seek relief from the condition on offering AIHS video services by filing a petition demonstrating that imposition of the condition no longer serves the public interest, convenience and necessity because there has been a material change in circumstance, including new evidence that renders the condition on offering AIHS video services no longer necessary in the public interest, convenience, and necessity." *Ibid.*, at paras. 122, 123, and 125, respectively.
[59] For a recent argument along these lines, see New.Net, "A Proposal to Introduce Market-Based Principles into Domain Name Governance" (31 May 2001), <http://www.icann.org/icp/icp-3-background/new-net-paper-31may01.pdf> and the following responses: ICANN, "Keeping the Internet a Reliable Global Public Resource: Response to New.net "Policy Paper"" (9 July 2001), <http://www.icann.org/icp/icp-3-background/response-to-new.net-09jul01.htm>, and K. Crispin, "Internet-draft: Alt-Roots, Alt-TLDs" (May 2001), <http://www.icann.org/stockholm/draft-crispin-alt-roots-tlds-00.txt>.

> bizarre as it may seem at first glance, the root server, and the various domain servers to which it points, constitute the very heart of the Internet, the Archimedean point on which this vast global network balances.

The need for coordination, in the interests of all network participants, is now rarely disputed. Indeed, these principles apply to any electronic communications network, as Andeen & King explain:[62]

> Ultimately, the fundamental technical driver of addressing is that Top Level Domains of any addressing scheme must be under the authority of a single, superordinate power if the network is to be globally effective. There is no way to avoid this.

What that authority might look like, and who should hold it, of course, are separate issues. We will return to them after considering some of the policy reasons for such authority in the context of all three of traditional telecommunications, the Internet, and ENUM.

### (b) ENUM Tiers 0 and 1 issues parallel Internet Tier 0 (root zone) issues

The technical and policy issues which arise at ENUM Tiers 0 and 1 have direct parallels in the Internet IP address and DNS Tier 0 zones. There are at least three groups of reasons why Tier 0-like resources are often subject to public oversight. The first is consumer protection. Preservation of personal privacy, security of communications, and prevention of 'identity theft' are but a few key consumer protection concerns which governments around the world must address (whether by legislation or 'supervised self-regulation'[63]). These considerations are perhaps even more applicable to ENUM because it aims to enable users to be reached on a number of different devices by means of just one number.[64] In many countries, there are also rules preventing the unauthorized switching of consumers' telephone services from one firm to another ('slamming') and the practice of charging customers for services which they have not ordered ('cramming').

Law enforcement is a second area. American telecommunications carriers, for example, are required to be able to provide access not only to traditional telecommunications traffic, but "packet-mode communications" as well, under the *Communications Assistance for Law Enforcement Act* (CALEA).[65] Any ENUM implementation in the U.S., for example, would almost certainly have to comply with these and other law enforcement-related requirements.

---

[60] See, for example, D.R. Johnson & D.G. Post, "And How Shall the Net Be Governed?: A Meditation on the Relative Virtues of Decentralized, Emergent Law," in B. Kahin & J.H. Keller, eds., *Coordinating the Internet* (Cambridge, MA: MIT Press, 1997) 62.
[61] D. Post, "Governing Cyberspace: 'Where is James Madison when you need him?'" (6 June 1999), <http://www.icannwatch.org/archive/governing_cyberspace.htm>.
[62] Note 27 above, at 251.
[63] M. Priest, "The Privatization of Regulation: Five Models of Self-regulation" (1997-98) 29 Ottawa L. Rev. 233.
[64] See J. Shiver, Jr., "Single-Number Plan Raises Privacy Fears" *Los Angeles Times* (2 September 2001), <http://www.latimes.com/news/printedition/la-000071061sep02.story>.

Third, to the extent that market competition is often the best guarantor of the public interest (particularly in areas such as price and service quality), it is important to recall that authoritative control of numbering resources is widely believed to be a *sine qua non* of telecommunications competition. This is not merely a transitional issue, either. It is one of the crucial tasks which modern regulators continue to perform indefinitely once competitive markets are achieved. This does not mean, however, that regulators themselves must perform all administrative and management functions. Rather, those are often delegated to independent third parties (such as industry consortia).[66] This is also how the E.164 country code "+1" is administered.[67] 'Local number portability' (LNP) is yet another justification for such control. As Andeen & King have remarked, "[t]here are many areas of uncertainty regarding the future of communications competition, but few are so fundamental in origin or far-reaching in implication as the rudimentary issue of addressing."[68] It is so fundamental that it is positively *required* in the countries which have signed on to the WTO Agreement on Basic Telecommunications and endorsed its associated "Reference Paper on Regulatory Principles"[69] (these countries together represent most of the world by traffic volume).

These basic public policy considerations (and arguably the Reference Paper itself), when added to the technical considerations discussed above, suggest that ENUM Tier 0 zone should be administered in an "objective, timely, transparent, and non-discriminatory manner." The dot at the far right of Internet domain names and email addresses, and the underlying IP address space itself, are both invisible to Internet users, but essential to Internet operation. For the exact same reasons as in traditional telecommunications and ENUM, *Internet* naming, numbering, and addressing resources should be administered in an objective, timely, transparent, and non-discriminatory manner – at the international level, since they are inherently international in scope.

---

[65] See 47 U.S.C. §1001, *et seq*. For background resources, see U.S. Department of Justice, Federal Bureau of Investigation, CALEA Implementation Section, "Ask CALEA" Website (undated), <http://www.askcalea.net/> and FCC, Third Report and Order, Communications Assistance for Law Enforcement Act, CC Docket 97-213 (released 31 August 1999), at para. 55.
[66] See notes 37 and 38 above, regarding telephone numbering administration and management in New Zealand and Canada, respectively.
[67] See note 21 above.
[68] Note 27 above, at 251.
[69] WTO Negotiating Group on Basic Telecommunications, "Reference Paper on Regulatory Principles" (24 April 1996), <http://www.wto.org/english/tratop_e/servte_e/tel23_e.htm>. Section 6, "Allocation of Scarce Resources", reads (in part): "Any procedures for the allocation and use of scarce resources, including frequencies, numbers and rights of way, will be carried out in an objective, timely, transparent and non-discriminatory manner."

### (c) ICANN

ICANN, for its part, appears to be aware of the important international policy role which it performs in the Internet Tier 0 environment. Its articles of incorporation say that:[70]

> …in recognition of the fact that the Internet is an international network of networks, owned by no single nation, individual or organization, the Corporation shall […] pursue the charitable and public purposes of lessening the burdens of government and promoting the global public interest in the operational stability of the Internet…

While the references to "charitable and public purposes" and "lessening the burdens of government" have more to do with U.S. tax law than lofty principles, the reference to the "global public interest" is powerful. ICANN's current president acknowledges this important duty in the most emphatic terms:[71]

> In linking the formation of ICANN to the global Internet community, the White Paper established a public trust that required that the DNS be administered in the public interest as the unique-rooted, authoritative database for domain names that provides a stable addressing system for use by the global Internet community. *This commitment to a unique and authoritative root is a key part of the broader public trust – to carry out the Internet's central coordination functions for the public good – that is ICANN's reason for existence*.

Yet even though it might appear that ICANN would have policy authority over whatever TLD is chosen for the ENUM root zone, it appears to either defer to the IAB on such matters, or to consider itself bound by an IETF-ICANN agreement[72] to treat all ENUM-related issues as *technical* issues, thus putting them in the IAB and IETF's bailiwick. The boundaries of the relationship between the IETF and ICANN are still being sorted out. The process that resulted in the creation of ICANN was driven primarily by the need to first institutionalize the DNS, and then add new gTLDs to it. The technical community was generally unmoved by these issues, and many in it viewed the desire by a few to mimic NSI's money-making machine as fundamentally 'un-Internet-like.' These individuals, often very influential in the Internet technical community, tried to ignore ICANN as much as possible.

It is somewhat puzzling that this community seems to wield so much more power than ICANN over the ENUM issue, which seems at least as policy-oriented as technical. Indeed, the issue of authority over top-level domains would seem to be squarely within ICANN's area of expertise, as it has spent the past three years painstakingly creating a structure of contracts binding registries and registrars in several of the other TLDs (including some of the ccTLDs, or country-code TLDs). Yet as this news story explains, ICANN has stayed out of ENUM so far:[73]

> Additionally, the Internet Corporation for Assigned Names and Numbers (ICANN), a nonprofit organization operating since 1998 by contract with the government, has some jurisdiction over the Internet's addressing system. It has so far declined to get involved, turning down proposals from private companies to set up their own systems under the ".tel" or ".num" domains. Given that the plan sits at the intersection of the telephone and the Internet realms, it presents a tricky policy issue for the United States.

---

[70] Section 3, Articles Of Incorporation Of Internet Corporation For Assigned Names And Numbers (As Revised November 21, 1998), <http://www.icann.org/general/articles.htm>.
[71] Note 26 above (reference omitted and emphasis added).
[72] RFC 2860, B. Carpenter, IAB; F. Baker, IETF; M. Roberts, ICANN, "Memorandum of Understanding Concerning the Technical Work of the Internet Assigned Numbers Authority" (June 2000), <http://www.ietf.org/rfc/rfc2860.txt>.
[73] A.E. Cha, ""Enum" Competition Pits Phone Industry Against Internet Start-Ups" *Washtech.com* (April 21, 2001), <http://www.washtech.com/news/netarch/9211-1.html>.

For its part, based on public consultation, the French numbering authority appears unwilling to enter into any ENUM-related agreements with ICANN, preferring to see the ITU hold whatever power ICANN might ordinarily hold over an ENUM TLD:[74]

> At the international level, a majority of contributions consider it desirable for the [ITU] to coordinate implementation of ENUM and to handle administration of the reference domain (Tier 0). The ICANN is considered to be an organization that is too young, too fragile, with no regulatory power, and too dependent on a single government to handle this coordination. The ITU, an entity that grows out of the United Nations organization, enjoys the benefit of years of experience with the rules for managing the international numbering system and can be a guarantor of neutrality. In this capacity, the ITU appears to be the best guarantor of consistency between E.164 numbers and ENUM domain names. Technical management of the domain could be handled by the organization designated by the ITU in concert with the ICANN.

Andrew McLaughlin, chief policy officer of ICANN, has been quoted as saying:[75]

> This is an area where over-regulation would be a tragedy. A lot of people look to ICANN to be an authority in this area, but we restrict our role very strongly. ENUM is not part of our mandate. […]
>
> If there isn't a balanced, fair, and open way of doing things, people won't use it and companies won't rely on it… Either ENUM is globally available to all users on equal terms or it's useless.

This statement would appear to suggest a surprising change of policy. Public Internet identifiers have always been subject to public oversight, varying from minimal (in the Postel era) to extensive (in the ICANN era), by the U.S. government. Indeed, if it were not for the DoC's heavy-handed intervention in the creation of ICANN, it is entirely possible that the DNS would now be essentially under the control of VeriSign/NSI. NSI fought tooth and nail the end of its registry and registrar monopolies, and due to the tremendously important role which it continues to play in the day-to-day operation of the Internet, it still enjoys prodigious bargaining power with the DoC. ICANN's then-chairperson Esther Dyson gave this evaluation of NSI's degree of cooperation during ICANN's early days:[76]

> Given this history, and the wealth that has been created through its administration of those government contracts, NSI is in no hurry to see that monopoly eroded. Since this very goal is a principal short-run objective of ICANN, NSI has apparently concluded that its interests are not consistent with ICANN's success. Thus it has been funding and otherwise encouraging a variety of individuals and entities to throw sand in the gears whenever possible, from as many directions as possible.

---

[74] France, Autorité de Régulation des Télécommunications (ART), "Principles and Conditions for Implementation of an ENUM Protocol in France – Abstract of Contributions to the Public Consultation" (English version) (July 2000), <http://www.art-telecom.fr/publications/synthese-enum-ang.htm>.
[75] In B. Michael, "The Politics of Naming" *Communications Convergence* (7 August 2001), <http://www.cconvergence.com/article/CTM20010710S0001>.
[76] ICANN, "Esther Dyson's Response to Questions," letter from E. Dyson to R. Nader and J. Love, Consumer Project on Technology (15 June 1999), <http://www.icann.org/chairman-response.htm>.

This is precisely the type of behaviour which economists expect from monopolies and dominant operators in network industries. To counteract it, either after the fact or from the outset, most liberalized nations enact rules to ensure that public communications networking identifiers are "globally available to all users on equal terms" (to use McLaughlin's words). That justification may also lead them to impose similar rules on ENUM, and the exact same considerations should be brought to bear with respect to Internet IP addresses and domain names.

An excellent 2000 article by Michael Froomkin lays to rest a number of myths about ICANN.[77] Froomkin demonstrates persuasively that: (*i*) ICANN is a creature of the U.S. Department of Commerce (DoC), not a 'self-organizing' emanation of the Internet community; (*ii*) ICANN is engaged in broad policy-making, not narrow technical coordination; (*iii*) ICANN is the *regulator* of the DNS and IP address space; (*iv*) most ICANN policy decisions are subject to the approval of the DoC; (*v*) the DoC has not given up any of its authority over the root zone, but rather only delegated limited administrative authority to ICANN, and (*vi*) the manner of that delegation likely violates one or both of U.S. administrative and constitutional law. He observes that: "[p]inning down the exact nature of DoC's relationship with ICANN is difficult, perhaps because a studied ambiguity on a few key points serves the interests of both parties."[78] This compared to the Internet technical community, where comparative legal documentation does not even exist! The following passage is Froomkin's most relevant for present purposes:[79]

> whichever characterization of the government's legal interest prevails, there is no dispute that the U.S. government, through the Department of Commerce, currently enjoys de facto control of the DNS. Nor is there any dispute that DoC has at least temporarily ceded to ICANN, through a variety of contractual and quasi-contractual agreements, almost all the control the United States enjoys. DoC has, however, explicitly reserved a right of review, the power to create new top-level domains, and the contractual right to replace ICANN with another body or take over DNS management directly.

The U.S. government has thus retained *residual policy authority* over the IP address space and DNS. This paper makes no comment about the manner of ICANN's performance of its delegated authority, but rather queries the proper locus of this residual policy authority, given the treatment of similar issues in the telecommunications and (possibly) ENUM worlds. If DoC's delegation of limited policy authority was indeed illegal under U.S. law, as Froomkin argues, then ART's characterization of ICANN as "fragile" would be quite accurate.[80]

The concern for many countries, of course, is that one country is effectively in control of a global network, even though significant authority has been delegated to an ostensibly 'international,'

---

[77] M. Froomkin, "Wrong Turn In Cyberspace: Using ICANN To Route Around The APA And The Constitution" (2000) 50 Duke L.J. 17, <http://personal.law.miami.edu/~froomkin/articles/icann.pdf>.
[78] *Ibid*., at 93.
[79] *Ibid*., at 166.
[80] Note 74 above.

'bottom-up,' 'private-sector' body.[81] As Froomkin observes: "DoC cannot quasi-privatize the DNS in a manner that allows the United States to retain ultimate control of the root zone file but achieve deniability about everything that its agent or delegate does with day-to-day control."[82] The desire to retain this control is certainly understandable, and even has precedent. As noted above, the nations sharing World Zone 1 have agreed to delegate certain administrative authority over it to the NANC, which in turn instructs the NANPA on its management. The NANC remains subject to the residual policy authority of the numbering plan administrators of member countries. Those countries, in turn, allocate administrative authority and management responsibilities for Numbering Plan Areas (NPAs, commonly known as area codes) within their territories. The important difference, of course, is that World Zone 1 is a *regional* resource, with joint regional governance. The DNS and IP address space, by contrast, are fundamentally *international* resources, as ICANN itself acknowledges.

### (d) The White Paper

The DoC process which culminated in the creation of ICANN was imposed in place of an ongoing, Internet community-based project which aimed to institutionalize the IANA function on an international footing. The International Ad Hoc Committee's[83] (IAHC) Final Report was animated by these basic principles:[84]

> The Internet top level domain space is a public resource and is subject to the public trust. Therefore any administration, use and/or evolution of the Internet TLD space is a public policy issue and should be carried out in an open and public manner in the interests and service of the public. Appropriately, related public policy needs to openly balance and represent the interests of the current and future stakeholders in the Internet name space.

For several reasons which are not relevant for present purposes,[85] the DoC unilaterally nullified the IAHC effort and declared in January 1998 in the Green Paper[86] that it would begin a rulemaking procedure to accomplish the corporatization and privatization of the IANA functions, according to the instructions of the White House in its "Framework for Global Electronic Commerce"[87] (which may go down in history as the high water mark of official belief in the unsuitability of government for governing). Not surprisingly, since it did not even refer to the IAHC process, the Green Paper was criticized by (mainly) non-American commenters for, among other things: slighting participation by the international community, being too US-centric, failing to recognize the need to implement an international approach, and interfering in a community consensus.[88] Froomkin believes that these comments were taken on board by the DoC:[89]

---

[81] Milton Mueller has demonstrated that ICANN, as presently constituted, is none of these things. See M. Mueller, "ICANN and Internet Governance: Sorting through the debris of self-regulation." (1999) 1 info 497, <http://www.icannwatch.org/archive/mueller_icann_and_internet_governance.pdf> (the assertions in which relating to the motivations and influence of trademark interests and ISOC in the creation of ICANN I am less comfortable with).
[82] Note 77 above, at 169.
[83] <http://www.iahc.org/>.
[84] IAHC, "Final Report of the International Ad Hoc Committee: Recommendations for Administration and Management of gTLDs" (4 February 1997), <http://www.gtld-mou.org/draft-iahc-recommend-00.html>.
[85] For a good description of these events, see C. Simon, "Overview of the DNS Controversy" (3 July 2001), <http://www.rkey.com/dns/overview.html>.
[86] United States Department of Commerce, "A Proposal to Improve Technical Management of Internet Names and Addresses: Discussion Draft 1/30/98" (30 January 1998) (the "Green Paper"), <http://www.ntia.doc.gov/ntiahome/domainname/dnsdrft.htm>.
[87] United States, The White House, "A Framework for Global Electronic Commerce" (1 July 1997), <http://www.ecommerce.gov/framewrk.htm>.

> The U.S. government's control over the DNS was more accidental than anything else, and U.S. officials were receptive to arguments by friendly governments that it was unreasonable for the United States to hold such power over a control point that seemed likely to be bound into the sinews of every developed economy's commercial, social, political, and even artistic life.

In response, the next version of the policy statement, the White Paper[90] (in which the DoC abandoned the idea of a proper rulemaking and 'punted' the most difficult issues to a private corporation to be named later) explicitly acknowledged the aims of the IAHC process, and the need for the proposed corporation to take an international approach in its work:[91]

> The U.S. Government believes that the Internet is a global medium and that its technical management should fully reflect the global diversity of Internet users. We recognize the need for and fully support mechanisms that would ensure international input into the management of the domain name system. In withdrawing the U.S. Government from DNS management and promoting the establishment of a new, non-governmental entity to manage Internet names and addresses, a key U.S. Government objective has been to ensure that the increasingly global Internet user community has a voice in decisions affecting the Internet's technical management.
>
> We believe this process has reflected our commitment. Many of the comments on the Green Paper were filed by foreign entities, including governments. Our dialogue has been open to all Internet users - foreign and domestic, government and private - during this process, and we will continue to consult with the international community as we begin to implement the transition plan outlined in this paper.

Perhaps most importantly for the purposes of this discussion, the White Paper also asserted that:[92]

> …the U.S. continues to believe, as do most commenters, that neither national governments acting as sovereigns nor intergovernmental organizations acting as representatives of governments should participate in management of Internet names and addresses.

The White Paper says that, despite best efforts to fix it, the gTLD-MoU process "was not able to overcome initial criticism of both the plan and the process by which the plan was developed,"[93] but soon after gives the real reason for its failure:[94]

> As a result of the pressure to change DNS management, and in order to facilitate its withdrawal from DNS management, the U.S. Government, through the Department of Commerce and NTIA [National Telecommunication and Information Administration], sought public comment on the direction of U.S. policy with respect to DNS, issuing the Green Paper on January 30, 1998.

---

[88] Excerpts from the comments of, respectively: America Online, Inc. (U.S.), Melbourne IT (Australia), the European Community, and Demon Internet (U.K.), summarized in J. Weinberg, "ICANN and the Problem of Legitimacy" (2000) 50 Duke L.J. 187 at note 103.
[89] Note 77 above, at 167.
[90] See note 2 above.
[91] *Ibid*. at Section 11.
[92] *Ibid*. at Section 4.
[93] *Ibid*. at "Background: The Need for Change".
[94] *Ibid*.

Two factors are thus said to have led the DoC to intervene: (*i*) pressure from some quarters to speed reform along, and (*ii*) the U.S. government's desire to "withdraw from DNS management." In retrospect, and notwithstanding the otherwise very high quality of drafting throughout the White Paper, definitions of the words 'withdraw' and 'management' would have been helpful. The White Paper states unequivocally that national governments (presumably including the U.S. government) should not 'participate' in management of Internet names and addresses. Yet perhaps the DoC never intended to withdraw from DNS administration or policy *entirely*. Ironically, of course, it would be three and a half more years before even the first instance of the top priority of reform – the addition of new TLDs – would happen.

### (e)      *International expectations*

The European Union was the most insistent on the international governance issue from the beginning. Its expectation, and that of other countries as well, of the U.S. government's *complete* withdrawal, has not been met. Rather, as a July 2000 report by the U.S. General Accounting Office (GAO) reveals, the DoC has "no current plans to transfer policy authority for the authoritative root server to ICANN, and therefore it has not developed a scenario or set of circumstances under which such control would be transferred."[95] Indeed, just the opposite has happened since then. Instead of being required to divest itself of either its registry or registrar business by May 10, 2001,[96] after a long and heated negotiation, NSI was allowed to maintain both businesses, under certain conditions, and continue in its role as registrar of the only commercially significant TLD, ".com", until November 10, 2007.[97]

In its July 1998 response to the White Paper, the European Commission appeared to accept that U.S.-based private sector management of Internet names and addresses was inevitable, but affirmed its belief in the importance of multilateral and intergovernmental arrangements:[98]

> It would consequently be appropriate for the EU to participate fully in encouraging the appropriate multilateral environment for the coordination of international policies in this area, including the necessary contribution of the international organisations. The international community can and should provide an appropriate political and legal framework for the future management of the Internet by the proposed private industry self-regulatory body, in the interests of its own stability.

---

[95] United States General Accounting Office, Office of the General Counsel, Letter to United States Senators and Congresspersons Re: Department of Commerce: Relationship with the Internet Corporation for Assigned Names and Numbers (7 July 2000) at 27, <http://www.gao.gov/new.items/og00033r.pdf>.
[96] As required in the November 1999 amendments to the Cooperative Agreement: see *Amendment 19 to the DOC/VeriSign Cooperative Agreement*, dated 10 November 1999, <http://www.icann.org/nsi/coopagmt-amend19-04nov99.htm>. The "Cooperative Agreement" began as *Network Information Services Manager(s) for NSFNET and the NREN: INTERNIC Registration Services Cooperative Agreement No. NCR-9218742 between National Science Foundation and Network Solutions, Incorporated* [sic], dated 1 January 1993, <http://www.ntia.doc.gov/ntiahome/domainname/nsi.htm>.
[97] See *Amendment 24 to the DOC/VeriSign Cooperative Agreement*, dated 25 May 2001, <http://www.ntia.doc.gov/ntiahome/domainname/agreements/amend24_52501.htm>.
[98] European Commission, "Communication from the European Commission to the European Parliament and to the Council: Internet Governance, Management of Internet Names and Addresses, Analysis and Assessment from the European Commission of the United States Department of Commerce White Paper" (29 July 1998), <http://europa.eu.int/ISPO/eif/InternetPoliciesSite/InternetGovernance/MainDocuments/com(1998)476.html>.

The US Government has also recognised that the Internet now has a major international dimension, an important step forward, which the EU can endorse and encourage. Such a realisation has not come lightly in certain US circles which still identify the Internet with US R&D programmes and US-based organisations.

The US White Paper has the merit of recognising that an US-centric approach is increasingly out-dated. Accordingly, there is now an opportunity for European and other international interests to take up the challenge to participate fully in the next phase of Internet development.

The opportunities for that participation have principally taken two forms. First, individual members of ICANN's board of directors might happen to be Europeans.[99] Second, the EU and European countries can participate voluntarily in ICANN's officially powerless Governmental Advisory Committee (GAC),[100] a body whose genesis is remarkably unclear considering the stance in the White Paper against the participation of national governments and intergovernmental organizations.[101] The government officials who attend GAC meetings clearly participate as representatives of sovereigns because they are sent there by them. Yet it must be an uneasy role, since the White Paper explicitly proscribes national governments from 'participating' in management of Internet names and addresses. As the myth of the Internet as ungovernable and even inherently resistant to law continues to wear away (and with it the political stigma of asserting the public interest in the Internet environment), national governments' interest in participating in Internet governance can be expected to increase. Whether the GAC provides them with an adequate forum to participate remains to be seen.

While not complaining too loudly about any one thing in particular, the European Commission has continually reaffirmed its expectation that the U.S. government withdraw completely from Internet governance. In an April 2000 report, the Commission notes:[102]

The broad scope of the powers and authorities reasserted by the US Administration (as recently as November 1999) notwithstanding, the US Department of Commerce has repeatedly reassured the Commission that it is still their intention to withdraw from the control of these Internet infrastructure functions and complete the transfer to ICANN by

---

[99] So long as they are not associated with any government. ICANN's Bylaws state, in part: "no official of a national government or a multinational entity established by treaty or other agreement between national governments may serve as a Director." Bylaws For Internet Corporation For Assigned Names And Numbers, A California Nonprofit Public Benefit Corporation, As Amended and Restated on October 29, 1999 and Amended Through July 16, 2000, <http://www.icann.org/general/bylaws.htm>, at Section 5.

[100] Its Website explains: "The Governmental Advisory Committee (GAC) of ICANN is an advisory committee comprising representatives of national governments, multinational government organisations and treaty organisations, and Distinct Economies as recognised in international fora. In accordance with the ICANN Bylaws the GAC's role is to "consider and provide advice on the activities of the Corporation as they relate to concerns of governments, particularly matters where there may be an interaction between the Corporation's policies and various laws, and international agreements." The GAC will operate as a forum for the discussion of government interests and concerns, including consumer interests. It is an advisory committee, and as such has no legal authority to act for ICANN, but will report its findings and recommendations to the ICANN Board." (<http://www.noie.gov.au/projects/international/DNS/gac/index.htm>).

[101] No Green or White Paper commenters called for the creation of the GAC. Nobody seems to be able to account for its appearance in ICANN's constating documents.

[102] European Commission, "Communication from the Commission to the Council and the European Parliament: The Organisation and Management of the Internet - International and European Policy Issues 1998 – 2000" (7 April 2000), <http://europa.eu.int/ISPO/eif/InternetPoliciesSite/Organisation/COM(2000)202.doc> at section 5.4.

October 2000. The Commission has confirmed to the US authorities that these remaining powers retained by the United States DoC regarding ICANN should be effectively divested, as foreseen in the US White Paper. The necessary governmental oversight of ICANN should be exercised on a multilateral basis, in the first instance through the Governmental Advisory Committee.

In a September 2000 resolution in response to this report, the Council noted (in part) that:[103]

> a number of important issues currently remain unresolved, in particular:
>
> (a) the nature of, and arrangements for, balanced and equal oversight of some of ICANN's activities by public authorities;
>
> […]
>
> (e) the transfer of the management of the root server system from the US Department of Commerce to ICANN, under appropriate international supervision by public authorities; [and noted]
>
> that those issues need to be addressed with due regard for both the interests of the international community as a whole and the public policy challenges involved, particularly as regards competition, personal data protection and respect for intellectual property rights; […]

The parallels between these issue sets and those implicated by ENUM are obvious. When the European Parliament responded to the Commission's April 2000 report, the Europeans' dismay and frustration with the lack of a truly international Internet governance solution was apparent. In the resolution, the Parliament says that (among other things) it:[104]

> insists that neither the Commission, nor the US Government, nor other governments should interfere in the organisation and management of the Internet, but they should give it sufficient independence and a legal basis at [the] international level, so that it may be an independent network […and…]
>
> [c]onsiders it necessary to guarantee the independence of ICANN from the US Government and to define the legal framework to which it must adhere in future, on the understanding that it is of paramount importance to maintain international neutrality if ICANN is to play a key role in the global development of the information society; […]

For its part, Canada, too, had high expectations for divestiture in the ICANN-forming period of 1998:[105]

> From the perspective of the Government of Canada, one of the most important goals of the reform process continues to be creation of a DNS coordinating body, the "new corporation," that will at a minimum be truly accountable and representative. *It is clearly not enough for the U.S. government to ensure merely that it has "privatized" the DNS -- i.e. divested U.S. government agencies of control of DNS functions and placed control in the hands of a "private sector" group.* The White Paper itself set a higher standard than this, and such bare-bones privatization will certainly not meet the needs of most end-user groups or of the international community.

---

[103] Council of the European Union, "Council Resolution on the organisation and management of the Internet" (28 September 2000) , <http://europa.eu.int/ISPO/eif/InternetPoliciesSite/InternetGovernance/ResolutionEN.html> at section 3.
[104] European Parliament, "European Parliament resolution on the Commission communication to the Council and the European Parliament on 'The Organisation and Management of the Internet – International and European Policy Issues 1998-2000'" (15 March 2001), <http://europa.eu.int/ISPO/eif/InternetPoliciesSite/InternetGovernance/EPResolution15March2001.html>.
[105] Canada, Industry Canada, "Reform of the Domain Name System: Current Developments & Statement of Principles: An Information Paper Prepared by Industry Canada with the assistance of Omnia Communications Inc." (September 1998), <http://e-com.ic.gc.ca/english/strat/651d22.html> at section 2.3 (emphasis added).

In late 1998, those countries which had expressed interest in helping to reform the management of Internet names and addresses[106] accepted the rather unorthodox idea of a California non-profit corporation managing most of the global public Internet's logical infrastructure. They so agreed in part because of the memes prevalent at the time that the Internet was uncontrollable and that government was incapable of addressing the public policy issues which it presented, but also because they were led to believe that private-sector *management* by the new corporation would eventually evolve into complete *administration* (i.e., policy authority), within – at the most – two years (a stabilizing transitional period). Instead, the U.S. government has retained so much residual policy authority over ICANN that Michael Froomkin concludes that:[107]

> as ICANN is utterly dependent on DoC for ICANN's continuing authority, funding, and, indeed, its reason for being, it would be reasonable to conclude that the corporation is currently so captive that all of ICANN's decisions can fairly be charged to the government.

Either a number of countries have seriously misinterpreted the White Paper for over three years, or the U.S. government has failed to meet their expectations for its complete withdrawal.[108] As Froomkin observed in late 2000, regarding the most basic questions of how to design a globally-effective governance structure for the Internet: "*[m]ost of these problems are if anything more real, and more pressing, today*."[109]

### 6. Unifying International Telecommunications Networks and Governance

In their excellent 1997 chapter on public network addressing and competition, Andeen & King speculate on the intertwined futures of the traditional telecommunications world and the Internet:[110]

> …one could argue that in the technical, administrative, and governance dimensions the Internet will absorb and subordinate the telephone world. As intriguing as this speculation might be, it is highly improbable that the Internet could subsume, much less handle, anything near the scale of the existing telephony infrastructure, particularly since that infrastructure carries most of the Internet's traffic. Nevertheless, this development suggests why the continuing blending of telephone and Internet form and functionality calls for a more penetrating assessment of the challenges of communications under competition. A focus on addressing provides a special window into the technical and socio-institutional problems at the heart of this transition.

This paper has attempted to provide an assessment of the challenges of pro-competitive telecommunications and Internet policy in the context of an emerging technology which in effect bridges the two. Ultimately, it should not matter whether ENUM is thought of as a telecommunications issue or an Internet issue because they engage the exact same global public interests. It is only due to historical reasons that they are presently subject to different governance patterns.

---

[106] There were not many of them – very few countries even knew what was happening.
[107] Note 77 above, at 27.
[108] Of course, another possibility is that the U.S. government does not consider that it is "acting as a sovereign" in the exercise of its residual policy authority over the DNS and IP address space. However, if it is not acting as a sovereign in its contractual relationship with VeriSign Global Registry Solutions, for example, then one wonders why VeriSign would feel the need to honour the Cooperative Agreement at all – particularly in not participating in efforts to establish 'alternative roots', in which it could operate free of regulation. *Black's Law Dictionary* (6th) (St. Paul, MN: West, 1990) defines "sovereign," in part, as a "person, body, or state in which independent and supreme authority is vested."
[109] Note 77 above, at 167 (emphasis added).

For the purposes of Internet legal and policy analysis, it is essential to distinguish the features and contingencies of the Internet's *logical infrastructure* layer from those of its *physical*, *application*, and *content/transaction* layers. Very different conditions and considerations prevail at each layer, and consequently legal arguments and policy approaches have to be tailored accordingly. In the logical layer, there is a technical need for uniqueness in naming, numbering, and addressing (among other things). This unavoidably creates a situation in which *someone* holds a certain degree of 'control' over them. As Weinberg explains, prior to ICANN, this control was always exercised in a 'public-regarding manner':[111]

> ICANN's task in seeking public acceptance of its legitimacy was made more complicated by the fact that it was a private entity seeking to play the sort of role more commonly played in our society by public entities. Its self-assigned task, after all, was one of setting rules for an international communications medium of surpassing importance. The task, administration of Internet identifiers, had historically been performed at the behest of the U.S. government and had been conducted in an explicitly public-regarding manner.

Although ICANN's current president acknowledges ICANN's obligation to continue this tradition for the benefit of Internet participants worldwide, residual policy authority over these resources continues to be held by the U.S. government. The U.S. has apparently created and then dashed expectations among other nations that it is committed to putting the governance of these resources onto a completely independent, completely international footing.

The internetworking paradigm is a tremendously powerful one. An open, accessible, non-proprietary global public network which offers universal interconnectivity and interoperability is a remarkable thing. In many ways, the Internet truly does turn previous models of communications networking on their heads, putting vastly more power into the hands of users to define services for themselves or invent services to offer to the public. The import of former IAB chair Brian Carpenter's simple but insightful observation that "nobody can turn it off"[112] cannot be overstated. In very large measure, there is no one standing astride the Internet who can say what anyone can or cannot do with it, on it, under it, or even around it (at least not yet – the imperial aspirations of Microsoft and AOL notwithstanding). Naming, numbering, and addressing resources are in fact a relatively narrow exception to this pattern – but an inordinately influential exception – which creates, in the words of David Post, an "Archimedean point on which this vast global network balances."[113] Aside from these unavoidable restraints, there is almost no other framework or opportunity for 'top-down' control of the Internet, from either a technical or a legal point of view. To many, this diffusion of power is the Internet's defining strength.

---

[110] Note 27 above, at 249-250.
[111] Note 31 above, at 215-216.
[112] RFC 1958, B. Carpenter, ed., "Architectural Principles of the Internet" (June 1996), <http://www.ietf.org/rfc/rfc1958.txt> at 4.
[113] Note 61 above.

Yet it is also the source of some of the public Internet's most worrisome weaknesses. The Internet community's struggle to control spam, denial-of-service attacks, and worms illustrates the downside of this diffusion of authority. Aside from downstream contractual obligations, there is currently no way to *require* network administrators to patch security holes in their publicly-accessible servers, and so the risks to overall Internet performance, and the costs of fighting these scourges, continue unabated. From an economic point of view, the commercial Internet industry continues to try to find a way to make sure that everybody gets paid for their contributions to overall interconnectivity (known in the telecommunications world as the 'settlement' process). Indeed, it is well-known that the Internet must be improved significantly, at the infrastructure layers, to meet the tremendous expectations which the world has for it: to be a ubiquitous industrial-grade, multiservice, multimedia communications platform. Rob Frieden describes this reality well:[114]

> An Internet-centric environment emphasizes the versatility of the Internet in terms of its ability to provide a medium for a wealth of different services and functions. But the Internet as we know it now will have to evolve and diversify, because a uniform, one-size-fits-all system cannot satisfy all particular user requirements. The Internet grows in importance because more users will resort to Internet-mediation for more services, including a variety of commercial applications. That will require Internet carriers and service providers to address and resolve a host of problems (e.g., quality of service, responding to consumer requirements, elasticity of demand-based pricing, customer care, network reliability, handling peak demand conditions) that perpetually have challenged telecommunications carriers.

Perhaps the most significant infrastructural challenge facing the Internet community today is the need to upgrade the Internet's basic protocol suite to IPv6.[115] The standard has been stable since at least 1998, yet for various reasons, it has not been widely implemented yet, despite the exhortations of ISOC. It so far appears that there is little interest in the Internet community (at least in North America) in making the required investment.[116] While it runs counter to Internet mythology, there was a time when the kind of 'top-down' policy authority which the DoC currently holds over the public Internet supported the imposition of such an upgrade. TCP/IP did not spontaneously become the sole basic protocol suite of the ARPANET. Bob Kahn, the co-inventor of TCP/IP, explains:[117]

---

[114] R. Frieden, *Managing Internet-Driven Change in International Telecommunications* (Boston: Artech House, 2001) at 131.
[115] RFC 2460, S. Deering, Cisco & R. Hinden, Nokia, "Internet Protocol, Version 6 (IPv6) Specification" (December 1998), <http://www.ietf.org/rfc/rfc2460.txt>.
[116] See J. Bound, "ISOC Member Briefing #4: IPv6 Deployment" (September 2001), <http://www.isoc.org/briefings/004/>. As Bound notes, "Different geographies will evolve at different rates. For example, because Europe and Asia have less IPv4 address space than North America, their need to evolve to the abundance of IPv6 address space is more time critical. North America appears to be very slow with their need to deploy IPv6 than those markets. But, the enterprise and telephony markets will require IPv6 for end-to-end transparency even in North America."
[117] R.E. Kahn, "The Role of Government in the Evolution of the Internet" (1994) Communications of the ACM, Vol. 37, No. 8, 15 at 17.

> The TCP/IP protocol adopted by DOD a few years earlier was only one of many standards. Although it was the only one that dealt explicitly with internetworking of packet networks, its use was not yet mandated on the ARPANET. However, on January 1, 1983, TCP/IP became the standard for the ARPANET, replacing the older host protocol known as NCP. This step was in preparation for the ARPANET-MILNET split, which was to occur about a year later. Mandating the use of TCP/IP on the ARPANET encouraged the addition of local area networks and also accelerated the growth in numbers of users and networks.

Kahn understates the phenomenon: standardization on TCP/IP triggered an *explosion* of interconnection. Hafner and Lyon describe the event this way:[118]

> As milestones go, the transition to TCP/IP was perhaps the most important event that would take place in the development of the Internet for years to come. After TCP/IP was installed, the network could branch anywhere; the protocols made the transmission of data from one network to another a trivial task.

It is still branching today. Brian Kahin and Bruce McConnell offer another view of the transition to TCP/IP and its significance:[119]

> A watershed decision during the mid-1980's was NSF's choice of the TCP/IP protocol rather than a proprietary protocol or X.25. As Mandelbaum and Mandelbaum observe:
>
>> "It led almost directly to the establishment of the system of specialized private academic networks we have today [in 1992], rather than to reliance by the academic and research community on the public, commercial networks that are the mainstays of the business world."

Both the design of TCP/IP and its adoption by the pre-commercial Internet community were fundamentally non-market phenomena. The Internet's core protocols and standards were initially developed for explicitly non-commercial purposes, and were only later adopted by commercial networks (many of which evolved directly out of the early regional institutional networks). This fundamental tension underlies many of the other challenges facing the Internet's infrastructure today, such as quality of service and security. IPv6 is expected to help both causes, but there is no agency which can "mandate" its implementation anymore.

Of course, there is no such agency in the telecommunications world which could mandate a global telephone industry software upgrade either. National authorities might require Internet operators to modify their systems for reasons of public policy, though, with or without the IETF's help (the example being CALEA[120]). On the subject of quality of service, it is interesting to note that most commercially significant IP telephony traffic travels over private, dedicated IP links – effectively 'bypassing' the public Internet as much as possible.[121] The foregoing are intended merely to identify the kinds of challenges which the Internet faces, and the limitations which the diffusion of authority over it imposes. As the Internet continues to evolve, it will be important for legal (and other) scholars to track the fate of the fundamental design principles that have made the Internet an open, accessible, non-proprietary global public network, offering universal interconnectivity and interoperability.[122] As Lessig has shown, there is nothing in its nature which ensures that it will always bear these remarkable characteristics. Rather, the economic characteristics of networks (e.g., network effects) and the ruthlessly binary nature of computer networks (you're either in or you're out, on or off) combine to make the question of who controls the Internet's code that much more important.

---

[118] K. Hafner & M. Lyon, *Where Wizards Stay Up Late: The Origins of the Internet* (New York: Simon & Schuster, 1996) at 249.
[119] B. Kahin & B. McConnell, "Towards a Public Metanetwork: Interconnection, Leveraging, and Privatization of Government-Funded Networks in the United States," in E. Noam & A. NiShuilleabhain, eds., *Private Networks Public Objectives* (Amsterdam: Elsevier Science, 1996) 307 at 315, quoting R. Mandelbaum & P.A. Mandelbaum, "The Strategic Future of the Mid-Level Networks," in B. Kahin, ed., *Building the Information Infrastructure* (New York: McGraw-Hill, 1992) 62 at note 6.
[120] See note 65 above. It is interesting to note that the IETF declined to include functionality designed to facilitate wiretapping in IETF standards-track documents because, among other things: "[t]he IETF, an international standards body, believes itself to be the wrong forum for designing protocol or equipment features that address needs arising from the laws of individual countries, because these laws vary widely across the areas that IETF standards are deployed in. Bodies whose scope of authority correspond to a single regime of jurisdiction are more appropriate for this task." (RFC 2804, IAB & IESG, "IETF Policy on Wiretapping" (May 2000), <http://www.ietf.org/rfc/rfc2804.txt> at 1). It is unclear whether this policy would restrict IETF's ability to develop protocols and features to enable ENUM at the Tier 1 and 2 (domestic) levels).

As in traditional telecommunications, and even the Internet itself, public oversight is necessary for those narrow (but incalculably important) aspects of ENUM relating to naming, numbering, and addressing. A single, coordinated global DNS domain for ENUM is called for at the Tier 0 level. Regardless of which TLD is ultimately chosen to host it, and how administrative and policy authority are divided up, the latter should reside at the international level. While this paper has not addressed Tier 1 issues as such, similar considerations likely call for a similar approach at the domestic level. However, that decision is up to each individual country. By contrast, since the Internet's logical infrastructure is inherently international, the kinds of public policy issues which arise at ENUM Tiers 0 and 1 arise in the Internet's own Tier 0 – the DNS root zone and IP address space. These issues should be thought of as matters of international public policy and dealt with in an objective, timely, transparent, and non-discriminatory manner at the international level. The existing pattern of Internet governance, and in particular the U.S. government's continuing residual policy authority over the DNS and IP address space, is not the outcome that many countries expected from the White Paper process. Internet governance is not yet truly international.

Private networks, to which the public does not ordinarily have access, do not raise such public policy issues, for obvious reasons. However, and contrary to the assertions of some,[123] there is very little that is *private* about the Internet's logical infrastructure.[124]  Indeed, the Internet may be the ultimate *public* network.[125]  At its physical layer, Internet-connected computers and telecommunications facilities are generally privately-owned and controlled (although the latter are often subject to public oversight where service is offered to the public).  At the application and content/transaction layers, there are practically no limits (other than those of generally-applicable laws) on the autonomous private action of Internet participants.  At the all-important logical infrastructure layer (which links them all together), however, things are very different.  The need for neutral, authoritative, international public oversight of shared elements of the Internet's logical infrastructure is succinctly described by Mark Lemley in the following statement, which serves well as a closing thought:[126]

---

[121] See note 8 above, at 6.
[122] A first-rate example of this type of work is M.A. Lemley & L. Lessig, "The End of End-to-End: Preserving the Architecture of the Internet in the Broadband Era" (2001) 48 UCLA L. Rev. 925 (an earlier version is available at: <http://www.bepress.com/blewp/default/vol2000/iss2/art8/>).
[123] Rutkowski asserts that "[n]either the Internet nor the WWW is a network or a service in the conventional sense - much less public ones. These and countless other systems, services, and activities are heterogeneous, autonomously self-organizing phenomena that do not require normal top-down institutions. They are NOT telecommunications." (A.M. Rutkowski, "International Institutional Roles and Values", presented at *The Impact of the Internet on Communications Policy*, Harvard University, Cambridge, MA (December 3-5, 1997), <http://www.ksg.harvard.edu/iip/iicompol/Papers/Rutkowski.html>).

> If the optimal number of Internets is indeed one, governance of the system itself must in the final analysis be effective at the global level.

---

[124] Just as ICANN performs (or procures the performance of) specific public services for the Internet community, the IETF is committed to 'open source' standards, and the crucial TCP/IP protocol suite, the *lingua franca* of the logical layer, is effectively in the public domain.  RFC 2026, S. Bradner, Harvard University, "The Internet Standards Process -- Revision 3" (October 1996), <http://www.ietf.org/rfc/rfc2026.txt> at 28: "In all matters of intellectual property rights and procedures, the intention is to benefit the Internet community and the public at large, while respecting the legitimate rights of others."

[125] This is the subject of C. McTaggart, "Governance of the Internet's Infrastructure: Network Policy for the Global Public Network" (1999) [unpublished Master of Laws (LL.M.) thesis, Faculty of Law, University of Toronto], <http://www.internetstudies.org/members/craimct/thesis/section0.htm>.

[126] M.A. Lemley, "The Law and Economics of Internet Norms," (1998) 73 Chi-Kent L. Rev. 1257 at 1282.